\documentstyle[aps,prb,psfig]{revtex} \begin{document}
\twocolumn[\hsize\textwidth\columnwidth\hsize\csname@twocolumnfalse\endcsname
\title{Surface Reconstructions and Bonding via the Electron Localization
Function: \\ The Case of Si(001)} 

\author{L. De Santis,$^{1,2}$ and R.  Resta$^{1,3}$} \address{ $^1$INFM --
Istituto Nazionale di Fisica della Materia\\ $^2$SISSA -- Scuola
Internazionale Superiore di Stud\^\i\ Avanzati, Via Beirut 4, 34014 Trieste,
Italy \\ $^3$Dipartimento di Fisica Teorica, Universit\`a di Trieste, Strada
Costiera 11, 34014 Trieste, Italy} 

\maketitle

\begin{abstract} The bonding pattern of a covalent semiconductor is disrupted
when a surface is cut while keeping a rigid (truncated bulk) geometry. The
covalent bonds are partly reformed (with a sizeable energy gain) when
reconstruction is allowed. We show that the ``electron localization
function'' (ELF)---applied within a first--principles pseudopotential
framework---provides un unprecedented insight into the bonding mechanisms. 
In the unreconstructed surface one detects a partly metallic character, which
disappears upon reconstruction. In the surface reformed bonds, the ELF
sharply visualizes strongly paired electrons, similar in character to those
of the bulk bonds. \end{abstract}

]
\narrowtext

\section{Introduction}

The driving mechanism for surface reconstructions of a covalent material is
the tendency of low--coordinated surface atoms to saturate dangling bonds. It
is then obvious that a completely different bonding pattern differentiates
the unreconstructed ({\it i.e.} bulk terminated) surface from the
reconstructed one. When several different reconstructions are compared, these
show in turn different bonding features. First--principles electronic
structure calculations have provided over the years a major insight into
these phenomena: however, the major tools used in investigating bonding at
large has been the analysis of either the electronic charge density or the
projected density of states. Since a few years a much more informative tool
has been introduced in the quantum chemical literature to deal in a
quantitative way with bonding features: this goes under the name of
``electron localization function'' (ELF).~\cite{Becke_90,Silvi_94,Savin_97}
Although bonding at a reconstructed surface looks like an ideal arena for an
ELF investigation, we are aware of only one such study, which however is
performed at the non--selfconsistent tight--binding level.~\cite{Fassler_95}
We present here a state-of-the-art thorough investigation about a
paradigmatic case: namely, the silicon (001) surface. We show that the
insight into the nature of bonding at the surface is very accurate and clear
when the ELF is analyzed. We demonstrate---through a series of contour plots
in high--symmetry planes---the outstanding ability of ELF in discriminating
between the surface bonds occurring upon dimerizations at the Si(001)
surface, thus providing an unprecedented insight into the physical mechanisms
which drive the reconstruction. Dangling orbitals and surface bonds are
visualized with a resolution incomparably sharper than by using the current
tools, such as charge--density plots or projected densities of states.

We are using here the approach which has become the ``standard model'' in
first--principle studies of covalent semiconductors:~\cite{Cohen91} namely,
density--functional theory (DFT)~\cite{HK,KS,DFT} with norm--conserving
pseudopotentials.~\cite{Pickett_89} This approach has got a dominant role in
the literature because of many of its key features, but ELF investigations
within it have been quite scarce so far: ELF has been originally proposed as
an all--electron tool, and as such has been mostly applied. Instead, we are
going to show in the present work that by getting rid of the core electrons
and focussing on the bonding electrons only, one makes the ELF message
particularly perspicuous. 

We study here an elemental system, where the key issue is metallic versus
covalent bonding.  Since the early days of electronic--structure theory, we
understand bonding in simple metals through the paradigm of pseudopotential
perturbation theory:~\cite{Harrison,pseudo} the valence electrons behave
basically as a free--electron gas in the region outside the ionic cores. At
the opposite extreme, we understand covalent bonding through the paradigm of
the H$_2$ molecule, whose electrons are strongly paired in a singlet state;
in a more general case, covalent bond is characterized by ``localization'' of
valence--electron pairs in appropriate regions of space. While
charge--density plots do not help much in discriminating metallic from
covalent, the ELF provides a quantitative measure of ``metallicity'' vs. 
``covalency'' of a given bond, or more generally of a given valence region of
the system. This measure is particularly significant in a pseudopotential
framework where---amongst other things---a simple metal immediately displays
the free--electron nature of its valence electrons. We furthermore observe
that ELF is by definition an orbital--free approach: this feature is
preserved when a first--principle framework (as opposed to a tight--binding
one) is adopted, as it is done here.

The case study chosen for this work is possibly the theoretically best known
semiconductor surface. Its most relevant features are therefore very well
understood in the literature: in particular, the unreconstructed surface is
metallic and the reconstructed one is insulating.~\cite{Dabrowski92} The aim
of the present paper is not to demonstrate anything at variance with the
common wisdom.  Instead, our aim is to show how the known features of this
surface can be recovered in a simple and meaningful way from an ELF analysis.
It is also worth to stress that the ELF {\it does not} make use of the
spectrum of the system, while instead is a pure ground state property. In
perspective, we are proposing ELF, in a modern pseudopotential
framework,~\cite{Cohen91,Pickett_89} as a powerful tool for investigating
bonding between over- or under--coordinated atoms in more complex, and worse
known, situations.

\section{Electron Localization Function}

ELF was originally introduced by Becke and Edgecombe as a measure of the
conditional probability of finding an electron in the neighborhood of another
electron with the same spin. Starting from the short--range behavior of the
parallel--spin pair probability, they defined a scalar function, conveniently
ranging from zero to one, that uniquely identifies regions of space where the
electrons are well localized, as occurs in bonding pairs or lone--electron
pairs. By definition, ELF is identically one either in any single--electron
wavefunction or in any two--electron singlet wavefunction: in both cases the
Pauli principle is ineffective, and the ground wavefunction is nodeless or,
loosely speaking, ``bosonic''. In a many--electron system ELF is close to one
in the regions where electrons are paired to form a covalent bond, while is
small in low--density regions; ELF is also  close to  one where the unpaired
lone electron of a dangling bond is localised. Furthermore, since in the
homogeneous electron gas ELF equals 0.5 at any density, values of this order in
inhomogeneous systems indicate regions where the bonding has a metallic
character.

A major advance is the ELF reformulation due to Savin {\it et
al.},~\cite{Savin_92} which is adopted in the present work. Owing to the Pauli
principle, the ground--state kinetic energy density of a system of fermions is
no smaller than the one of a system of bosons at the same density: ELF can be
equivalently expressed in terms of the Pauli excess energy density, thus making
no explicit reference to the pair distribution. This is particularly convenient
in our case, since the pair density is outside the scope of DFT. The Savin {\it
et al.} reformulation also provides a meaningful physical interpretation: where
ELF is close to its upper bound, electrons are strongly paired and the electron
distribution has a local ``bosonic'' character. 

The definition of ELF is: \begin{equation} \label{e:ELF} {\cal E}({\bf
r}) = \frac{1}{1+[D({\bf r})/D_{h}({\bf r})]^2}, \end{equation} thus
taking by design values between zero and one. Following Savin {\it et al.},
$D({\bf r})$ is the Pauli excess kinetic energy density,
defined as the difference between the kinetic energy density and the
so--called von Weizs\"{a}cker kinetic energy functional:~\cite{Dreizler}
\begin{equation} D({\bf r})=\frac 12\nabla _{{\bf r}}\nabla _{{\bf
r^{\prime }}}\rho ({\bf r},{\bf r^{\prime }})\bigg |_{{\bf r}={\bf
r^{\prime }}}-\frac 18\frac{|\nabla n({\bf r})|^2}{n({\bf r})}
\label{e:D} \end{equation} where $\rho $ is the one--body reduced
(spin--integrated) density matrix. The von Weizs\"{a}cker functional
provides a rigorous lower bound for the exact kinetic energy
density\cite{Dreizler} and is ordinarily indicated as the ``bosonic''
kinetic energy, since it coincides with the ground--state kinetic energy
density of a non--interacting system of bosons at density $n({\bf r})$.
Therefore, $D$ is positive semidefinite and provides a direct measure of
the local effect of the Pauli principle. The other ingredient of
Eq.~(\ref{e:ELF}) is $D_{h}({\bf r})$, defined as the kinetic energy
density of the homogeneous electron gas at a density equal to the local
density: \begin{equation} D_h({\bf r})=\frac 3{10}(3\pi ^2)^{\frac
23}n({\bf r})^{\frac 53}. \label{e:D_h} \end{equation} It is obvious to
verify that the function ${\cal E}$ is identical to one in the ground
state of any two--electron system, while is identical to 0.5 in the
homogeneous electron gas at any density.

As commonly done in many circumstances---including in other ELF
investigations\cite{Savin_92}---we approximate the kinetic energy of the
interacting electron system with the one of the noninteracting Kohn--Sham
(KS) one. We therefore use the KS density matrix: \begin{equation} \rho ({\bf
r},{\bf r^{\prime }})= 2 \sum_i\phi_i^{KS} ({\bf r})\phi _i^{*KS}( {\bf
r^{\prime }}), \label{e:rho} \end{equation} where $\phi _i^{KS}({\bf r})$ are
the occupied KS orbitals. Such approximation is expected to become
significantly inaccurate only in the case of highly correlated materials.

The original ELF definition is an all--electron one, and has the remarkable
feature of naturally revealing the entire shell structure for heavy atoms. 
Such a feature is of no interest here, since only one electronic shell (the
$sp$ valence one) is involved in the bonding of Si atoms in any
circumstances. The pseudopotential scheme simplifies the landscape, since
only the electrons of the relevant valence shell are dealt with explicitly:
the ELF message comes out therefore much clearer, with basically no loss of
information.\cite{Kohout_97} In the spatial regions occupied by core
electrons, the pseudo--electronic distribution shows a depletion and ELF
assumes very low values. Ouside the ionic cores, in the regions relevant to
chemical bonding, the norm conservation endows the pseudocharge density with
physical meaning, as widely discussed in the modern pseudopotential
literature:~\cite{Cohen91,Pickett_89} in the specific case of Si, there is an
experience of about 20 years in dealing with a plethora of different physical
problems. By the same token of norm conservation, even the
pseudo--orbitals---and hence their kinetic energy density---closely map the
all--electron ones in the bonding regions. The pseudo ELF carries
therefore---in the material of interest here---the same information as the
all--electron ELF, while it removes irrelevant and confusing features due to
the inner, chemically inert, shells.

\section{Calculations}

The $(001)$ silicon surface is strongly reconstructed, with several different
reconstructed structures. A relatively simple $(2\times 1)$ reconstruction is
formed by surface atoms moving together in pairs to form dimers. 
\begin{figure}
\centerline{\psfig{figure=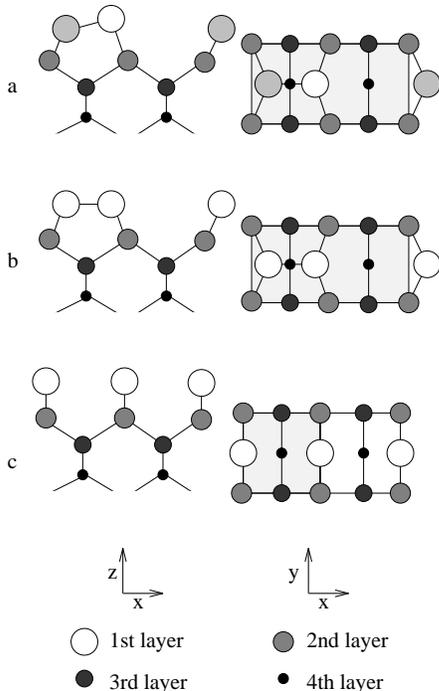}}
\caption{Side (left) and top (right) views of the $(2 \times 1)$
reconstructions of the Si(001) surface. (a) Buckled--dimer, (b)
symmetric--dimer, and (c) ideal (bulk--terminated) surface. The atoms in the first four
layers are shown, with symbols as indicated in the figure bottom.
The surface unit cells (shaded areas) are also shown.}
\label{f:Fig1}
\end{figure}
The driving force for the dimer formation is the elimination of a dangling bond from the
surface atoms. Each atom of the unreconstructed surface is bonded to only two
neighbours and therefore has two dangling bonds projecting out of the surface.
If the atoms move together in pairs, forming a new bond between them, then one
of these dangling bonds will be eliminated from each member of the pair. This
leads to a considerable energy gain since half of the broken surface bonds are
reconstructed. The buckled dimer model is a simple modification of the dimer
model in which each dimer is tilted or ``buckled'' out of the plane of the
surface. We have considered therefore the unreconstructed and the $(2\times 1)$
symmetric and buckled dimer surfaces, in order to better understand the
connection between surface reconstruction and electronic localization. A
schematic side and top view of these three surfaces is shown in
Fig.~{\ref{f:Fig1}.

\begin{figure}
\centerline{\psfig{figure=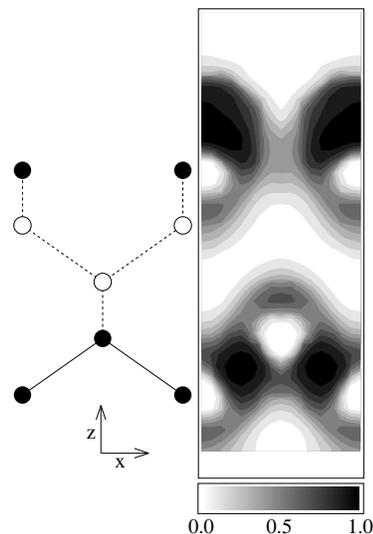}}
\caption{ELF contour plot for the $(001)$ bulk terminated silicon surface,
in the plane containing two nearest--neighbor atoms: the
relative dangling orbitals are revealed by the black ELF regions. The
grey--scale is also shown: dark (clear) regions correspond to large
(small)  ELF values. In the schematic side view, black balls and solid
lines correspond, respectively, to in--plane atoms and bonds; white balls
and dashed lines to off--plane atoms and bonds.}
\label{f:Fig2}
\end{figure}
All the calculation use quite standard ingredients: a plane-wave expansion of
the KS orbitals with a $10$ Ry kinetic--energy cut--off, $16$ {\bf k}-points on
a Monkhorst and Pack~\cite{MP} mesh for the irreducible Brillouin zone
integration, and a norm--conserving pseudopotential in fully non--local
form.\cite{Kleinman_82} All our calculations were performed in a supercell
geometry where the surface is modeled by a finite--size slab periodically
repeated in the direction normal to the surface. For the ideal (bulk 
terminated)
surface we use a unit cell containing $11$ layers of atoms separated by the
equivalent of $13$ atomic layers of vacuum. For the symmetric and buckled dimer
$(2\times1)$ surface, we have reproduced the Roberts and
Needs~\cite{Roberts_90} calculation, using directly their relaxed atomic
coordinates.
\begin{figure}
\centerline{\psfig{figure=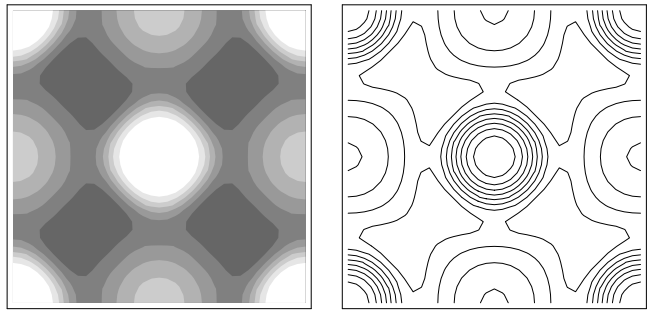}}
\caption{Contour plots for bulk aluminum in the [100] cristalline plane. 
Left panel: ELF, where the same grey--scale as in all ELF figures is adopted. 
Right panel: pseudocharge density.}
\label{f:Fig3}
\end{figure}

\begin{figure}
\centerline{\psfig{figure=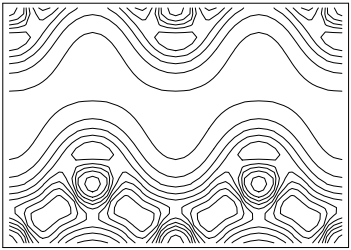}}
\caption{Pseudocharge density contour plot for bulk silicon in the
same crystallographic plane as for Fig.~\protect\ref{f:Fig2}.}
\label{f:Fig4}
\end{figure}
We show in Fig.~\ref{f:Fig2} the ELF contour plot for the unreconstructed
surface, in the plane containing the dangling orbitals and two nearest
neighbor surface atoms. As already discussed above, pseudoatoms
correspond to the white regions (ELF minima). Before discussing the surface
results, we pause to illustrate the bulk features and the capability of ELF
in discriminating between metallic and covalent bonding.

\begin{figure}[b]
\centerline{\psfig{figure=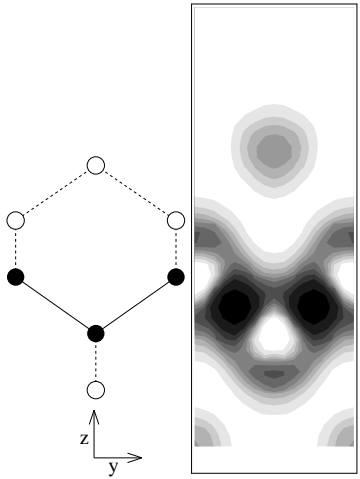}}
\caption{ELF contour plot for the $(001)$ bulk terminated silicon surface.
Plane orthogonal to the dangling bond, passing through the center of the
surface unit cell. For the schematic side view, symbols as in Fig.~\protect\ref
{f:Fig2}.}
\label{f:Fig5}
\end{figure}
In the bulk Si region, one clearly sees the almost black regions in
Fig.~\ref{f:Fig2} between nearest--neighbor atoms, whose bonding pattern in
this geometry has the shape of a ``zig--zag''chain. In these bonding regions
our calculated ELF attains the maximum value of 0.96, thus indicating that
the Pauli principle has little effect. In agreement with a chemical picture
of the covalent bond, we associate these regions to the opposite--spin
electron pairs---actually a ``bosonic'' system---localized between every pair
of bonded atoms.  This is to be contrasted to metallic bonding, where the
valence electrons have a free--electron nature: to make this point better
clear, we take as an example a paradigmatic simple metal: crystalline
aluminum, whose bulk ELF is shown in the left panel of Fig.~\ref{f:Fig3} to the
same grey scale. Comparing this plot with the bulk region of silicon in
Fig.~\ref{f:Fig2} we immediately appreciate the spectacular ELF ability to
distinguish in a very clearcut way between metallic bonding and covalent
bonding.  The ELF plot in aluminum shows---outside the core regions---a large
grey area, which correspond to a jellium--like (or Thomas--Fermi) ELF value. 
Actually, the maximum value attained by ${\mathcal E}({\bf r})$ between
nearest---neighbor atoms is only 0.61.

Comparison of the two ELF plots provides therefore the most significant and
perspicuous visualization of the important {\it qualitative} difference
between the covalent bond and the metallic one. In the bulk of the two
materials the Pauli principle plays quite a different role. The other typical
tools for analysis, such as charge--density plots or projected density of
states, lack by far a similar sharpness. As an example, we show the
corresponding pseudocharge density plots in the right panel of Fig.~\ref{f:Fig3}
(aluminum) and in Fig.~\ref{f:Fig4} (silicon). From such figures, hardly any
information about metallicity or covalency can be drawn.

\section{Results}

We now move to discuss the surface results. Focussing on the surface region
in Fig.~\ref{f:Fig2}, one notices that the two dangling orbitals are also
well visualized by the ELF plot: every dangling bond is associated to an {\it
isolated} electron, thus corresponding to an high ELF value. However, a more
striking bonding feature of the same unreconstructed Si(001) surface is
detected when analyzing ELF in the orthogonal plane, passing midway between
two nearest neighbor surface atoms (Fig.~\ref{f:Fig5}). There is a grey
``fermionic channel''---defined as a region where ELF is of the order of
0.5---between the two nonbonded atoms: this channel in the midpoint has a
perfect circular section, but it is actually formed by two grey--ELF strips
going round the surface atoms and intersecting at the midpoint, as clearly
shown by a top view (left panel of Fig.~\ref{f:Fig6}).
\begin{figure}
\centerline{\psfig{figure=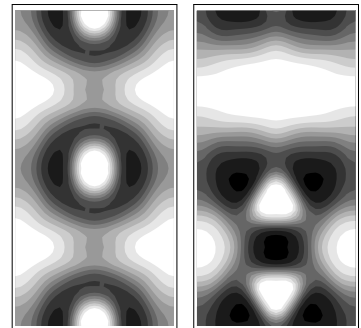}}
\caption{ELF contour plot for the Si$(001)$ bulk terminated (left panel) and
symmetric--dimer reconstructed (right panel). Surface plane passing through
the topmost atoms.}
\label{f:Fig6}
\end{figure}

The presence of this extended metallic--like system along the unrelaxed
surface must be connected to the strong surface strain, due to the nonbonded
atoms. As soon as the surface is relaxed and the dimer formation is allowed,
there is a strong surface ``bosonization'', associated with a sizeable
reduction of the fermionic channel. This localization effect due to the
dimerization process is perspicuously shown in Fig.~\ref{f:Fig6}, where the
unreconstructed situation is compared to the reconstructed one, for the
symmetric--dimer case. The metallic character of the unreconstructed surface,
and the insulating character of the reconstructed one, are well documented in
the literature.~\cite{Dabrowski92} What is remarkable here is that ELF
visualizes such characters without using {\it any} spectral information about
the system.
\begin{figure}
\centerline{\psfig{figure=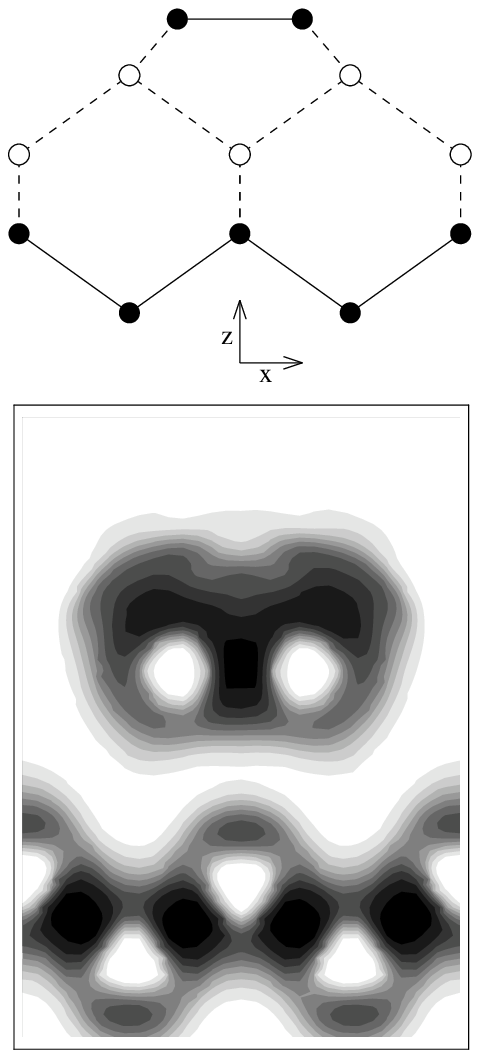}}
\caption{ELF contour plot for the Si$(001)$ $(2\times1)$ symmetric--dimer
surface. Plane orthogonal to the surface, containing the dimer. For the
schematic side view, symbols as in Fig.~\protect\ref{f:Fig2}.}
\label{f:Fig7}
\end{figure}

A more detailed analysis of the ELF variations induced by rebonding at the
surface is provided by Fig.~\ref{f:Fig7} for the symmetric--dimer case and in
Fig.~\ref{f:Fig8} for the buckled--dimer one, both drawn in the plane passing
through the surface atoms, and to be compared to the analogous plot of
Fig.~\ref{f:Fig2} for the unreconstructed case. In both cases the dimer has a
strongly covalent bond character and is surrounded by a region of high ELF
values: such localization effect is clearly connected to the surface reduced
coordination, since the ELF around the isolated topmost atoms has a more
pronounced atomic character.\cite{note}
\begin{figure}
\centerline{\psfig{figure=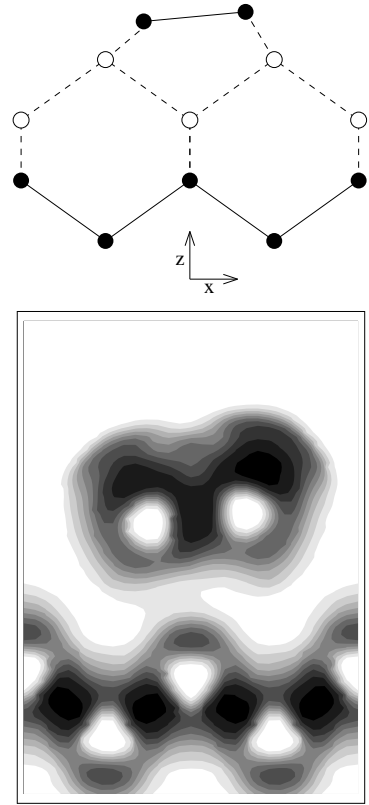}}
\caption{ELF contour plot for the Si$(001)$ $(2\times 1)$ buckled dimer
surface. Plane orthogonal to the surface, containing the dimer. For the
schematic side view, symbols as in Fig.~\protect\ref{f:Fig2}.}
\label{f:Fig8}
\end{figure}

In conclusion, we have shown how the electronic system at a paradigmatic
semiconductor surface changes its character upon reconstruction. ELF clearly
visualizes how electron pairing accompanies rebonding. In the unreconstructed
surface, besides the dangling--bond electrons, we have detected channels of
metallic--like character in the surface plane. Both features disappear upon
reconstruction.


\begin{references}

\bibitem{Becke_90}  A. D. Becke, K. E. Edgecombe, J. Chem. Phys. {\bf 92},
5397 (1990).

\bibitem{Silvi_94}  B. Silvi, A. Savin, Nature {\bf 371}, 683 (1994).

\bibitem{Savin_97}  A. Savin, R. Nesper, S. Wengert and T. F. F\"{a}ssler,
Angew. Chem. Int. Ed. Engl. {\bf 36}, 1808 (1997).

\bibitem{Fassler_95} T. F. F\"assler, U. H\"aussermann and R. Nesper, Chem.
Eur. J. {\bf 1}, 625 (1995).

\bibitem{Cohen91} M.L. Cohen, in {\it Electronic Materials}, edited by J.R.
Chelikowsky and A. Franciosi (Springer, Berlin, 1991), p.57.

\bibitem{HK}  P. Hohenberg and W. Kohn, Phys. Rev. {\bf 136} B864 (1964).

\bibitem{KS}  W. Kohn and L. J. Sham, Phys. Rev. {\bf 140}, A1133 (1965).

\bibitem{DFT} {\it Theory of the Inhomogeneous Electron Gas}, edited
by S.  Lundqvist and N.H. March (Plenum, New York, 1983).

\bibitem{Pickett_89}  W. E. Pickett, Comput. Phys. Rep. {\bf 9}, 115 (1989).

\bibitem{Harrison} W.A. Harrison, {\it Pseudopotentials in the Theory of
Metals} (Benjamin, New York, 1966).

\bibitem{pseudo} V. Heine, in {\it Solid State Physics}, edited by H.
Ehrenreich, F. Seitz, and D. Turnbull, vol {\bf 24} (Academic, New York, 1970),
p.1; M.L. Cohen and V. Heine, {\it ibid.} p.250.

\bibitem{Dabrowski92} J. Dabrowski and M. Scheffler, Appl. Surf. Sci. {\bf
56--58}, 15 (1992).

\bibitem{Savin_92}  A. Savin, O. Jepsen, J. Flad, O. K. Andersen, H. Preuss
and H. G. von Schneiring, Angew. Chem. Int. Ed. Engl. {\bf 31}, 187 (1992).

\bibitem{Dreizler}  R.M. Dreizler and E.K.U. Gross, {\it Density Functional
Theory: an Approach to the Quantum Many-body Problem} (Springer, Berlin,
1990).

\bibitem{Kohout_97}  M. Kohout and A. Savin, J. Comp. Chem. {\bf 18}, 1431
(1997).

\bibitem{MP}  H. J. Monkhorst and J. D. Pack, Phys. Rev. B {\bf 13}, 5188
(1976).

\bibitem{Kleinman_82} L. Kleinman and D. M. Bylander, Phys. Rev. Lett. {\bf
48}, 1425 (1982).

\bibitem{Roberts_90}  N. Roberts and R. J. Needs, Surf. Sci. {\bf 236}, 112
(1990).

\bibitem{note}  In a free pseudoatom, ELF has spherical symmetry with a single
maximum in the valence--shell region.

\end{references}
\end{document}